\documentclass[prl,twocolumn,superscriptaddress,letterpaper,amssymb,amsmath,floatfix]{revtex4-1}
\pdfoutput=1

\usepackage{amsmath}
\usepackage{stmaryrd}
\usepackage{color}
\usepackage{graphicx}
\usepackage{mathrsfs}

\begin{document}

\title{A Possible Fermi liquid in the lightly doped Kitaev spin liquid}

\author{Jia-Wei Mei}
\affiliation{Institute for Theoretical Physics, ETH Z\"urich, 8093 Z\"urich, Switzerland}

\date{\today}

\begin{abstract}
  In this paper, we study the lightly doped Kitaev spin liquid (DKSL) and find it to be a Fermi liquid state. The DKSL has a closed Fermi surface around the center of the Brillouin zone with the quantized volume determined by the Luttinger's theorem.  It has well-defined quasiparticles near the Fermi surface and can be detected in the angle-resolved photoemission spectroscopy. The DKSL has the topological Kitaev spin liquid (KSL) as its neutral background. It violates the Wiedemann-Franz law and has a large Wilson ration. These results has the potential experimental test in the iridates upon doping.  
\end{abstract}

\maketitle

Doping a spin liquid brings new phases in the strongly correlated electron systems and is believed to be the key physics in the high $T_c$ cuprates\cite{Lee2006}. Actually, the parent compounds of the cuprates are antiferromagnetic ordered, not in a spin liquid state. Many efforts have been made to find the spin liquid state in the quantum frustration Mott insulators\cite{Powell2011}. In Ref. \onlinecite{Kitaev2006}, Kitaev proposed an exactly solvable spin model with the spin liquid ground state on the honeycomb lattice. Recently the Kitaev model has the potential experimental realization in the strong spin orbit coupling magnets, such as the layered iridates A$_2$IrO$_3$ (A=Na, Li)\cite{Singh2010,Singh2011}. In these materials, it is proposed in Refs. \onlinecite{Jackeli2009,Chaloupka2010} that the spin interactions include both the isotropic Heisenberg term and the anisotropic Kitaev term. The Kitaev spin liquid (KSL) phase is stable when the Heisenberg term is small\cite{Chaloupka2010,Jiang2011}.  The exact KSL state and its experimental realization provide us the unique theoretical opportunity to test the physics of the doped spin liquid. This may give us some hints in the understanding of the high $T_c$ cuprates.

The doped Kitaev spin liquid (DKSL) was studied in Ref.\onlinecite{You2011} by using the SU(2) slave boson method\cite{Wen1996,Lee2006} and in Ref.\onlinecite{Hyart2011} by using the U(1) slave boson method\cite{Zou1988}. Both of them obtained the $p$-wave superconductors  upon doping. Motivated by these studies, we will make another investigation into the conducting state in the DKSL. In contrast to the $p$-wave superconductors in Refs.\onlinecite{You2011,Hyart2011}, we find a Fermi liquid state in this paper. Before doping, the KSL  has the topological order; it has four-fold degeneracy on the torus. The topological order is protected by the $Z_2$ gauge structure and is robust against any local perturbations\cite{Wen2002}. The KSL is a resonating valence bond (RVB) state\cite{Anderson1987, Burnell2011} and the DKSL is relevant to high $T_c$ cuprates in the broader context of the doped RVB state\cite{Anderson1987,Lee2006}.  In Refs. \onlinecite{You2011,Hyart2011},the topological robustness of the KSL breaks down after doping. None of the slave methods for the electron decomposition satisfies the $Z_2$ gauge structure. To protect the $Z_2$ gauge symmetry upon doping, we will use the dopon representation of the electron operators\cite{Ribeiro2005,Ribeiro2006} to study the DKSL. The dopon theory is a full fermionic decomposition of the physical electron operator. It describes the DKSL in terms of two different fermionic components: ``spinons'', the neutral spin-1/2 excitations of the KSL; ``dopons'', which describes the dopant holes and has the charge $e$ and spin-1/2. In the lightly DKSL, the hybridizing between spinons and dopons vanishes on the mean field level. The dopons form a Fermi liquid surrounded by the background spinons. The DKSL has well-defined low energy quasiparticles on the Fermi surface enclosing the quantized volume determined by the Luttinger's theorem\cite{Luttinger1960}. The Fermi surface is electron like regardless of whether we dope the holes or electrons into the KSL. These properties can be tested in the angle-resolved photoemission spectroscopy (ARPES) experiments. The background neutral spinons also have physical observable contributions. So unlike the standard Landau Fermi liquid\cite{Abrikosov1963}, the DKSL has a temperature-dependent specific heat coefficient and violates the Wiedemann-Franz law in the transport measurements.  In the T=0K limit, the DKSL has a large Wilson ratio. 

\begin{figure}
  \begin{center}
	\includegraphics[width=6cm]{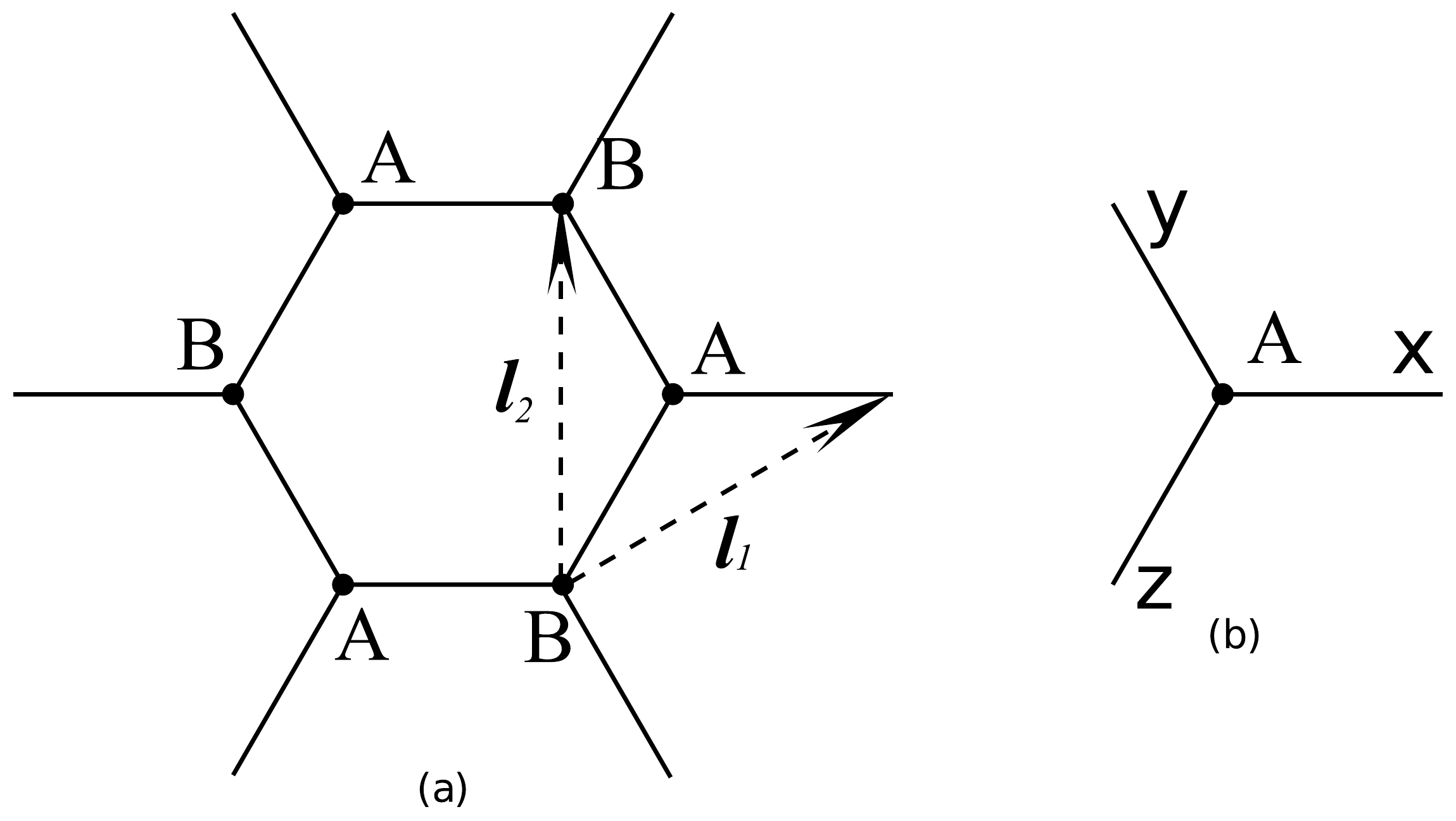}
  \end{center}
  \caption{The honeycomb lattice and the $x$, $y$ and $z$ links joining $A$ and $B$ sublattices. $\mathbf{l}_1=\frac{\sqrt{3}}{2}\hat{\mathbf{e}}_1+\frac{1}{2}\hat{\mathbf{e}}_2$ and $\mathbf{l}_2=\hat{\mathbf{e}}_2$ are the primitive vectors.}
  \label{fig:honeycomb}
\end{figure}
The KSL has the parent Hamiltonian of the spin $S=1/2$ system on the honeycomb lattice\cite{Kitaev2006}:
\begin{eqnarray}
  H_K=-J\sum_{\alpha\text{-links}}S_m^\alpha S_n^\alpha,\quad \alpha=x,y,z \label{eq:kitaev_model}
\end{eqnarray}
where $S_m^\alpha$ is $\alpha$-component of the $S=1/2$ spin operator on the site $m$ and the summation runs over all the nearest neighbor $\alpha$-links  ($\alpha=x,y,z$) joining A and B sublattices oriented in the $\alpha$-th direction as shown in Fig. \ref{fig:honeycomb}. The primitive vectors in the honeycomb lattice can be chosen as $\mathbf{l}_1=\frac{\sqrt{3}}{2}\hat{\mathbf{e}}_1+\frac{1}{2}\hat{\mathbf{e}}_2$ and $\mathbf{l}_2=\hat{\mathbf{e}}_2$.

To exactly solve the spin model (\ref{eq:kitaev_model}), Kitaev introduced four Majorana fermions $b_m^x$, $b_m^y$, $b_m^z$ and  $c_m$ to rewrite the spin operators as $S_m^\alpha=ib_m^\alpha c_m$ $(\alpha=x,y,z)$ with the constraint on every site $D_m=b_m^xb_m^yb_m^zc_m=\frac{1}{4}$. In this paper, the Majorana fermions are normalized such as $\{\gamma_m^\alpha,\gamma_n^\beta\}=\delta_{mn}\delta_{\alpha\beta}$ with $\gamma_m^{x,y,z}=b_m^{x,y,z}, \gamma_m^0=c_m$. In terms of the Majorana fermions, the Kitaev model (\ref{eq:kitaev_model}) now has the form 
\begin{eqnarray}
  H_K=J\sum_{\alpha\text{-links}}ic_mc_n\hat{u}_{mn}^\alpha, \quad\alpha=x,y,z
  \label{eq:km_mf}
\end{eqnarray}
with $\hat{u}_{mn}^\alpha=ic_m^\alpha c_n^\alpha$ and $\hat{u}_{mn}^\alpha=-\hat{u}_{nm}^\alpha$. Without loss of generality, we assume conventionally that $m$ is in A sublattice.  Using the four Majorana fermions, we can also construct the fermionic Schwinger representation for the spin operators\cite{Burnell2011}:
\begin{eqnarray}
  F_m&=&\begin{pmatrix}f_{m\uparrow} &f_{m\downarrow}^\dag\\ f_{m\downarrow} &-f_{m\uparrow}^\dag\end{pmatrix}
	=\frac{1}{\sqrt{2}}\sum_{\alpha=x,y,z,0}\gamma_m^\alpha \sigma^\alpha\\
	S_m^\alpha&=&-\frac{1}{4}\text{tr}[( F_mF_m^\dag-I )\sigma^\alpha], \alpha=x,y,z
  \label{eq:F_i}
\end{eqnarray}
where $\sigma^\alpha (\alpha=x,y,z)$ are the Pauli matrices and $\sigma^0=i I$. We have the single-occupancy constraint for the Schwinger fermions $G_m^\alpha=-\frac{1}{4}\text{tr}[(F_m^\dag F_m-I)\sigma^\alpha]=0$. The mapping between the Majorana fermions and the Schwinger fermions is not unique and all such mappings are equivalent under SU(2) gauge transformations.  The key observation of Kitaev is  $[H,\hat{u}_{mn}^\alpha]=0$ and $\hat{u}_{mn}^\alpha$ are the constants of motion with eigenvalues $U_{mn}^\alpha=\pm1/2$.  The KSL is a RVB state with the parent mean field Hamiltonian\cite{Burnell2011} 
\begin{eqnarray}
  H_K=&&J\sum_{\alpha-\text{links}}U_{mn}^\alpha \hat{u}_{mn}^0-2E_\text{vp}\sum_{x\text{-links}}U_{mn}^x \hat{u}_{mn}^x\nonumber\\
 &&-2E_\text{vp}\sum_{y\text{-links}}U_{mn}^y \hat{u}_{mn}^y -2E_\text{vp}\sum_{z\text{-links}}U_{mn}^z \hat{u}_{mn}^z 
  \label{eq:kmsf}
\end{eqnarray}
with $\hat{u}_{mn}^0=\frac{i}{2}(f_{m\uparrow}^\dag f_{n\uparrow}-f_{m\uparrow}^\dag f_{n\uparrow}^\dag)+\text{h.c.}$, $\hat{u}_{mn}^x=i(f_{m\downarrow}^\dag f_{n\downarrow}+f_{m\downarrow}^\dag f_{n\downarrow}^\dag)+\text{h.c.}$, $\hat{u}_{mn}^y=\frac{i}{2}(f_{m\downarrow}^\dag f_{n\downarrow}-f_{m\downarrow}^\dag f_{n\downarrow}^\dag)+\text{h.c.}$ and $\hat{u}_{mn}^z=\frac{i}{2}(f_{m\uparrow}^\dag f_{n\uparrow}+f_{m\uparrow}^\dag f_{n\uparrow}^\dag)\nonumber+\text{h.c.}$ Here $E_\text{vp}$ is the energy of the nearest neighbor vortex-pair on the honeycomb lattice.\cite{Baskaran2007}  The Hamiltonian (\ref{eq:kmsf}) is invariant under local $Z_2$ transformations 
\begin{eqnarray}
  f_{m\sigma}\rightarrow\tilde{f}_{m\sigma}=G_mf_{m\sigma},~{U}_{mn}^\alpha\rightarrow{\tilde{U}}_{mn}^\alpha=G_m{U}_{mn}^\alpha G_n
  \label{eq:Z2}
\end{eqnarray}
where $G_m$ is an arbitrary function with only the two values $\pm1$. Such a $Z_2$ gauge symmetry leads to the topological order with four-fold degeneracy for the KSL on the torus. This four-fold degeneracy is protected by the $Z_2$ gauge structure and is robust against any local perturbations\cite{Wen2002}. The wavefunciton of the KSL is a projection of the ground state of the parent Hamiltonian (\ref{eq:kmsf}), $|\text{KSL}\rangle=\mathcal{P}|\Psi\rangle_{\text{MF}}$ ($\mathcal{P}$ removes the double occupancy). Such a wavefunciton of the KSL is an exact result after the projection.   

Upon doping, we study the doped Kitaev model $H=H_t+H_K$ with the doping level $x$ on every site. $H_t$ is the hopping Hamiltonian upon doping
\begin{eqnarray}
  H_t=-t\sum_{\langle mn\rangle\sigma}\mathcal{P}c_{m\sigma}^\dag c_{n\sigma}\mathcal{P}
  \label{eq:dkm}
\end{eqnarray}
We will employ the dopon representations for the electron operators\cite{Ribeiro2005,Ribeiro2006}
\begin{eqnarray}
  c_{m\sigma}^\dag=\frac{1}{\sqrt{2}}\mathcal{P}_df_{m\sigma}^\dag\left( \sum_{\sigma'}\sigma' f_{m\sigma'}d_{m-\sigma'} \right)\mathcal{P}_d
  \label{eq:dopon_rep}
\end{eqnarray}
Here the spinon $f_{m\sigma}$ and the dopon $d_{m\sigma}$ are both fermionic operators. On every site, the states $|\uparrow0\rangle$, $|\downarrow0\rangle$, and the local singlet state $\frac{1}{2}(|\uparrow\downarrow\rangle-|\downarrow\uparrow\rangle)$ are the physical states mapping onto the states $|\uparrow\rangle$, $|\downarrow\rangle$ and the vacancy state $|0\rangle$, respectively. The operator $\mathcal{P}_d$ is to project out the unphysical triplet states between the spinon and the dopon on every site. The spinon sector is always half-filled on the honeycomb lattice, $\sum_\sigma f_{m\sigma}^\dag f_{m\sigma}=1$.

Here we slack our steps to give some explanations for such a dopon decompositions (\ref{eq:dopon_rep}). The dopon theory describes the DKSL in terms of two fermions: ``spinons'', the neutral spin-1/2 excitations of the KSL; ``dopons'', which has the same charge $e$ and spin-1/2 as the dopant holes. Here we start from the Mott insulator at half-filling and described the DKSL as a doped Mott insulator. 

The spinons in the dopon theory are always at half-filling on the honeycomb lattice. The Mottness is always described by the half-filled spinons even upon doping. The KSL at half-filling has the $Z_2$ gauge symmetry (\ref{eq:Z2}) and the nontrivial topological order. The topological order plays very important roles in spin liquids\cite{Wen2002}  and robust against local perturbations. However, the topological robustness is lost in the slave boson methods used in Refs. \onlinecite{You2011,Hyart2011} even in the very lightly doped case. Because the physical electron operator $c_{i\sigma}$ is no longer invariant under the $Z_2$ gauge transformation (\ref{eq:Z2}). This is very unacceptable if we have the simple belief in the topological order. Fortunately, the dopon theory still has the $Z_2$ gauge symmetry (\ref{eq:Z2}) upon doping. In the dopon representation, the electron operator has two spinon $f_{m\sigma}$ multipliers and remains unchanged under the $Z_2$ gauge transformations.    

The dopons in the dopon theory are fermions different from the slave boson methods in  Refs. \onlinecite{You2011,Hyart2011}. As noted in Ref \onlinecite{You2011}, Willans \textit{et al} showed in Ref. \onlinecite{Willans2010} that a spin vacancy with a $\pi$-flux is stable  as the low energy excitation in the dilute limit vacancy doped KSL. The doping process introduces the fermionic low energy excitations for the charged dopant holes. So it is reasonable for  us to treat the dopons as fermions in the lightly DKSL.

The dopons have the spin-1/2 degree of freedom which is another vital difference from the slave boson methods in  Refs. \onlinecite{You2011,Hyart2011}. The spinons are always at half-filling. So the spin degree of freedom in the dopon sector is used to neutralize the spin on the spin vacancy. Ref. \onlinecite{Willans2010} showed that the dopant holes have the low excitations as a spin vacancy with a $\pi$-flux. The flux in the KSL has something to do with the spin excitations described by $\hat{u}_{mn}^\alpha$ in Eq. (\ref{eq:km_mf}).\cite{Baskaran2007} The spin-1/2 degree of freedom in dopon sector is consistent with the numerical results.\cite{Willans2010}

Honestly, there is no rigorous proof for the validity of the dopon theory; nevertheless, it quite makes sense for us  to use it in the lightly DKSL.  We rewrite the hopping Hamiltonian as 
\begin{eqnarray}
  H_t&=&-\frac{t}{8}\sum_{\langle mn\rangle}\mathcal{P}_d[\text{tr}(\sigma^zD_mD_n^\dag)-\text{tr}(D_m\sigma^z(F_mF_m^\dag-I)\nonumber\\
  &&\times\sigma^zD_n^\dag)-\text{tr}(D_m\sigma^z(F_nF_n^\dag-I)\sigma^zD_n^\dag)+\text{tr}(D_m\sigma^z\nonumber\\
  &&\times(F_mF_m^\dag-I)(F_nF_n^\dag-I)\sigma^zD_n^\dag\sigma^z)]\mathcal{P}_d
  \label{eq:hopping_term}
\end{eqnarray}
with the definition $D_i\equiv\begin{pmatrix}d_{m\uparrow} & d_{m\downarrow}\\ d_{m\downarrow}^\dag &-d_{m\uparrow}^\dag\end{pmatrix}$. The process of hybridizing spinons and dopons, $B_m\propto\langle \text{tr}(D_m F_m)\rangle\neq0$, breaks both the $Z_2$ gauge symmetry and the physical U(1) electromagnetic symmetry\cite{Ribeiro2005,Ribeiro2006} resulting in the $p$-wave superconductors similar to the results in Refs. \onlinecite{You2011,Hyart2011} .  However, the KSL has a gap $E_\text{vp}$ in the spectrum. The spontaneous mixing $B_m$ competes with the gap energy $E_\text{vp}$ and is not allow for lightly doped case $x<\frac{E_{\text{vp}}}{t/2}\simeq0.134$. In this paper, we set $t=J=1$ and $E_\text{vp}=0.0668J$.\cite{Kitaev2006} 

Without the mixing between the dopons and spinons, $B_m\propto\langle \text{tr}(D_m F_m)\rangle=0$, the four-operator and six-operator terms in Eq. (\ref{eq:hopping_term}) are decoupled into bilinear terms such as $D_m D_n$ and $F_m F_n$ on the mean field level. Therefore, the spinons and dopons are decoupled on the mean field level. The DKSL is described as the Fermi liquid formed by the dopons surrounded in the KSL formed by spinons. The full effective mean field Hamiltonian is now
\begin{eqnarray}
  H_\text{MF}=H_d+H_f
  \label{eq:fullmf}
\end{eqnarray}
Here the dopon Hamiltonian reads out
\begin{eqnarray}
  H_d=\sum_{\mathscr{A}\mathbf{k}\sigma}\epsilon_{\mathscr{A}\mathbf{k}}^dd_{\mathscr{A}\mathbf{k}\sigma}^\dag d_{\mathscr{A}\mathbf{k}\sigma}+2xN\mu_d
  \label{eq:sector}
\end{eqnarray}
with $\mathscr{A}=1,2$ and $\epsilon_{1,2\mathbf{k}}^d=\pm\frac{t}{2}|e^{i\mathbf{k}\cdot\mathbf{l}_1}+e^{i\mathbf{k}\cdot\mathbf{l}_2}+1|-\mu_d$. $N$ is the number of the unit cells. The chemical potential $\mu_d$ is determined by the particle number of the dopant holes $x$. For $x=0.1$, $\mu_d=-1.324$. The spinon Hamiltonian is now ($U_{mn}^\alpha=\frac{1}{2}$)
\begin{eqnarray}
  H_f=\sideset{}{'}\sum_{\mathbf{k}}(\psi_{\mathbf{k}\uparrow}^\dag M_{\mathbf{k}\uparrow}\psi_{\mathbf{k}\uparrow}+\psi_{\mathbf{k}\downarrow}^\dag M_{\mathbf{k}\downarrow}\psi_{\mathbf{k}\downarrow})
    \label{eq:mfm}
\end{eqnarray}
Here $\sideset{}{'}\sum_{\mathbf{k}}$ takes the summation over half the Brillouin zone. $\psi_{\mathbf{k}\sigma}^\dag=\begin{pmatrix}f_{1\mathbf{k}\sigma}^\dag & f_{2\mathbf{k}\sigma}^\dag & f_{1-\mathbf{k}\sigma} & f_{2-\mathbf{k}\sigma}\end{pmatrix},	\sigma=\uparrow/\downarrow$, is the Nambu representation. The Hamiltonian matrices $M_{\mathbf{k}\sigma}$ are now 
\begin{eqnarray}
  M_{\mathbf{k}\uparrow}&=&\begin{pmatrix}0 & i g_-(\mathbf{k}) & 0 &-ig_+(\mathbf{k})\\-ig_-(-\mathbf{k}) &0 &ig_+(-\mathbf{k})&0\\0 & -i g_+(\mathbf{k}) &0 &ig_-(\mathbf{k})\\ig_+(-\mathbf{k}) &0 &-ig_-(-\mathbf{k}) &0 
  \end{pmatrix}\nonumber\\
  M_{\mathbf{k}\downarrow}&=&\begin{pmatrix}0&-ih_+(\mathbf{k})&0&-ih_-(\mathbf{k})\\ih_+(-\mathbf{k})&0&ih_-(-\mathbf{k})&0\\0&-ih_-(\mathbf{k})&0&-ih_+(\mathbf{k})\\ih_-(-\mathbf{k})&0&ih_+(-\mathbf{k})&0\end{pmatrix}
  \label{eq:Mm}
\end{eqnarray}
Here we have the definition $g_{\pm}(\mathbf{k})=\frac{J_\text{eff}}{4}(e^{-i\mathbf{k}\cdot\mathbf{l}_1}+e^{-i\mathbf{k}\cdot\mathbf{l}_2}+1)\pm \frac{E_\text{vp}^\text{eff}}{2}$ and $h_\pm(\mathbf{k})=\frac{E_\text{vp}^\text{eff}}{2}(e^{-i\mathbf{k}\cdot\mathbf{l}_1}\pm e^{-i\mathbf{k}\cdot\mathbf{l}_2})$ with the effective constants $J_\text{eff}=J[(1-x)^2-xt/2]$ and $E_{\text{vp}}^{\text{eff}}=E_{\text{vp}}[(1-x)^2-xt/2]$ . The spinon Hamiltonian has the eigenvalues $\pm\epsilon_{1\mathbf{k}\uparrow}^f$, $\pm\epsilon_{2\mathbf{k}\uparrow}^f$, $\pm\epsilon_{1\mathbf{k}\downarrow}^f$ and $\pm\epsilon_{1\mathbf{k}\downarrow}^f$ with the definition $\epsilon_{1\mathbf{k}\uparrow}^f=\frac{J_\text{eff}}{2}|e^{i\mathbf{k}\cdot\mathbf{l}_1}+e^{i\mathbf{k}\cdot
  \mathbf{l}_2}+1|$ and $\epsilon_{2\mathbf{k}\uparrow}^f=\epsilon_{1\mathbf{k}\downarrow}^f=\epsilon_{2\mathbf{k}\downarrow}^f=E_\text{vp}^\text{eff}$.
\begin{figure}
  \includegraphics[width=3cm]{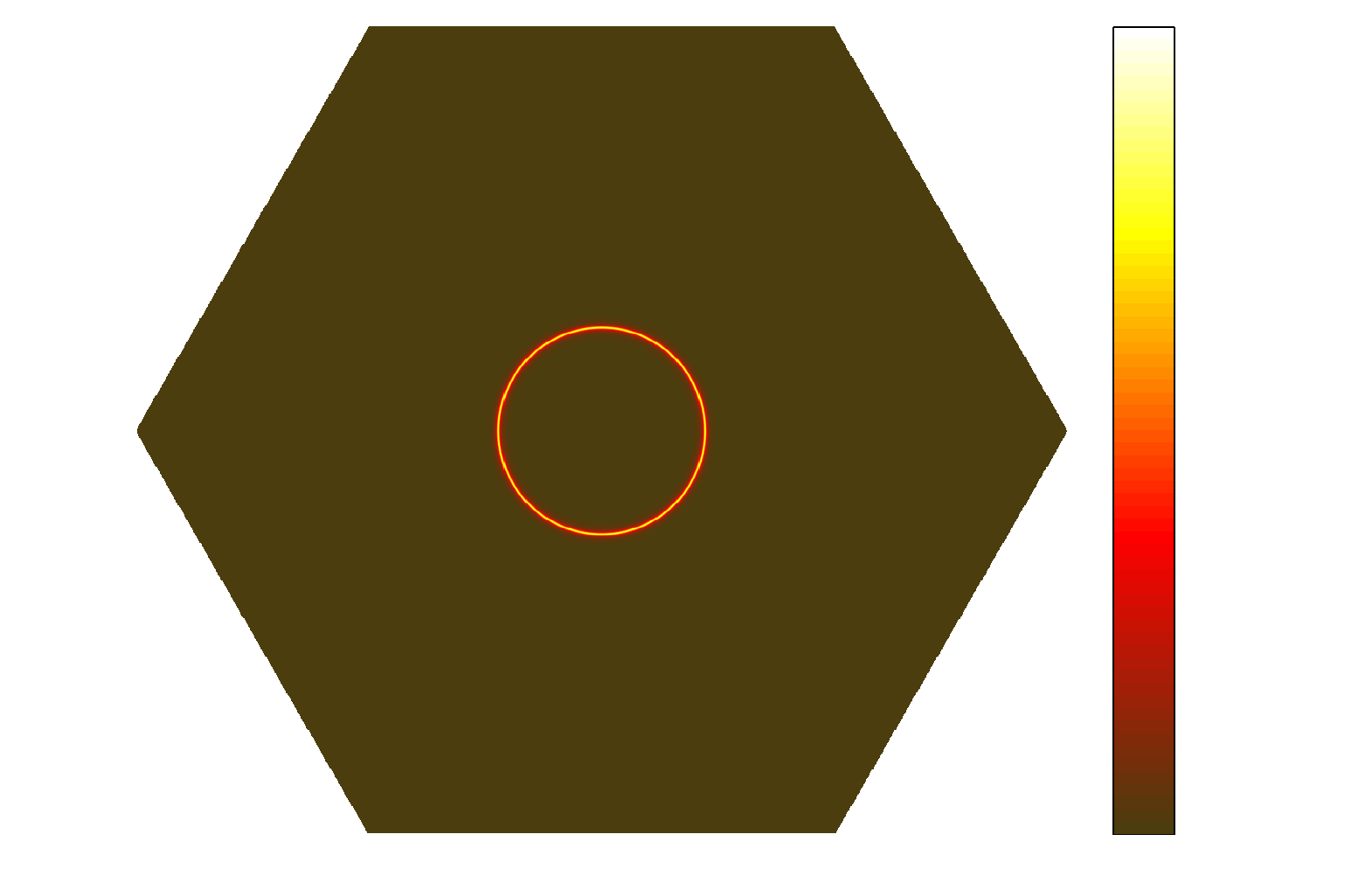}\includegraphics[width=6cm]{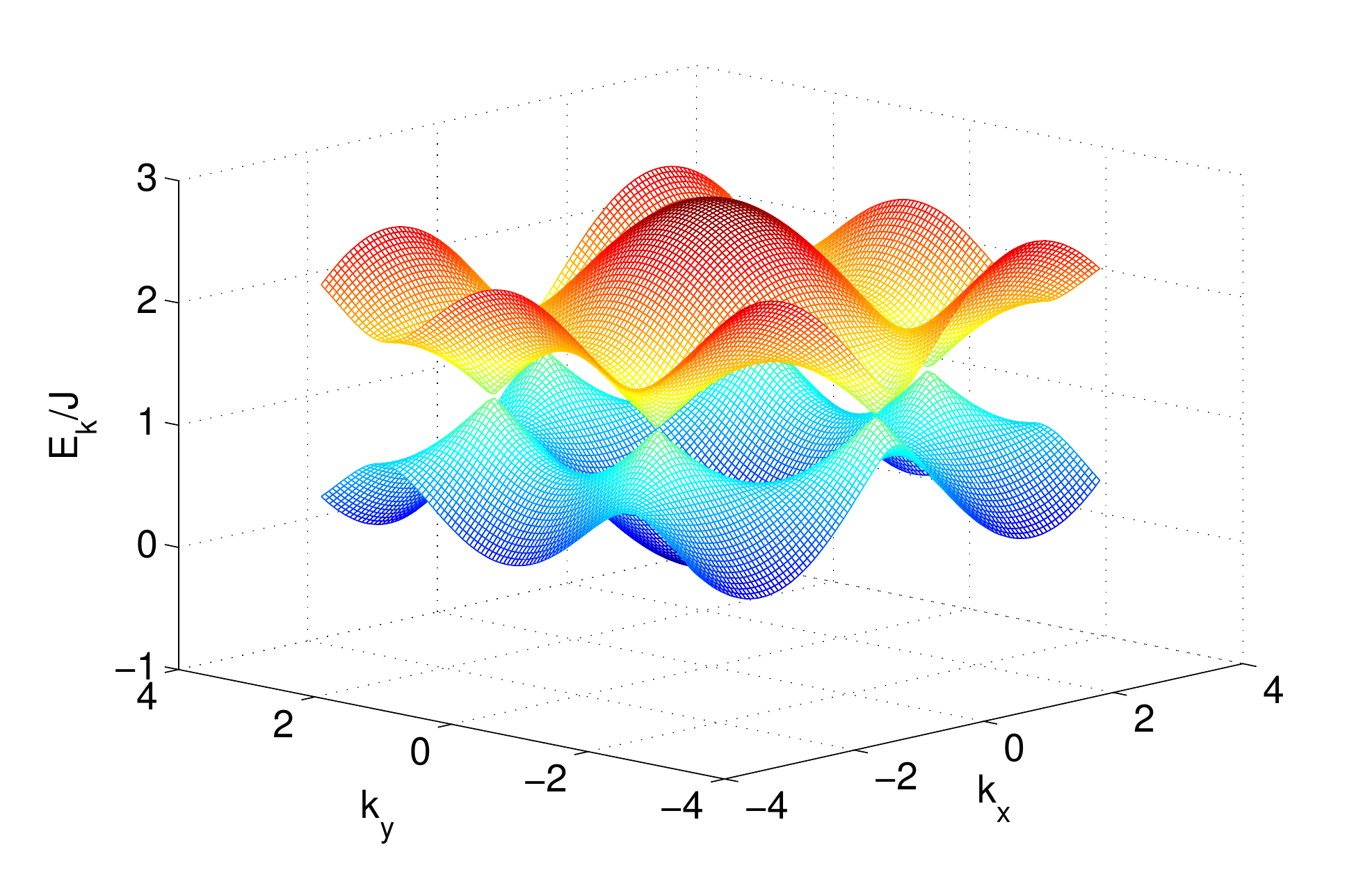}
  \caption{[Color online] (a) The ARPES intensity $I(\mathbf{k},0)$ near the Fermi energy for the DKSL at $x=0.1$. (b) The dispersions in the ARPES measurements. }
  \label{fig:arpes}
\end{figure}

In the dopon representation, we can write down the electron Green function $G_c$ in the DKSL
\begin{eqnarray}
  G_c(\mathbf{k},\omega)&\propto& G_d(\mathbf{k},\omega)+G_\text{inc}\nonumber\\
  &=&\sum_\mathscr{A}\frac{1}{\omega+i\gamma-\epsilon_{\mathscr{A}\mathbf{k}}^d}+\cdots
  \label{eq:greenfunction}
\end{eqnarray}
where $G_d(\mathbf{k},\omega)$ is the Green function for the dopon sector and $G_\text{inc}$ is the incoherent part. The dopon sector forms a Fermi liquid and the real part of the electron Green's function $\text{Re}G_c$ satisfies the Luttinger's theorem\cite{Luttinger1960,Abrikosov1963}
\begin{eqnarray}
  \frac{1}{N}\sum_{\mathbf{k}\sigma}\text{Re} G_c(\mathbf{k},0)>0=2x
  \label{eq:lth}
\end{eqnarray}
$\text{Re} G_c(\mathbf{k},\omega)$ is the real part of the electron Green function. The image part of the electron Green function is proportional to the intensity $I(\mathbf{k},\omega)\propto\text{Im}G_c(\mathbf{k},\omega)$ in the ARPES measurements. The intensity at Fermi energy is shown in Fig. \ref{fig:arpes}(a). The dispersion is shown in Fig. \ref{fig:arpes}(b). Regardless of whether we dope the holes or electrons into the KSL, the Fermi surface is always electron-like pockets around the Brillouin center.

\begin{figure}
  \includegraphics[width=6cm]{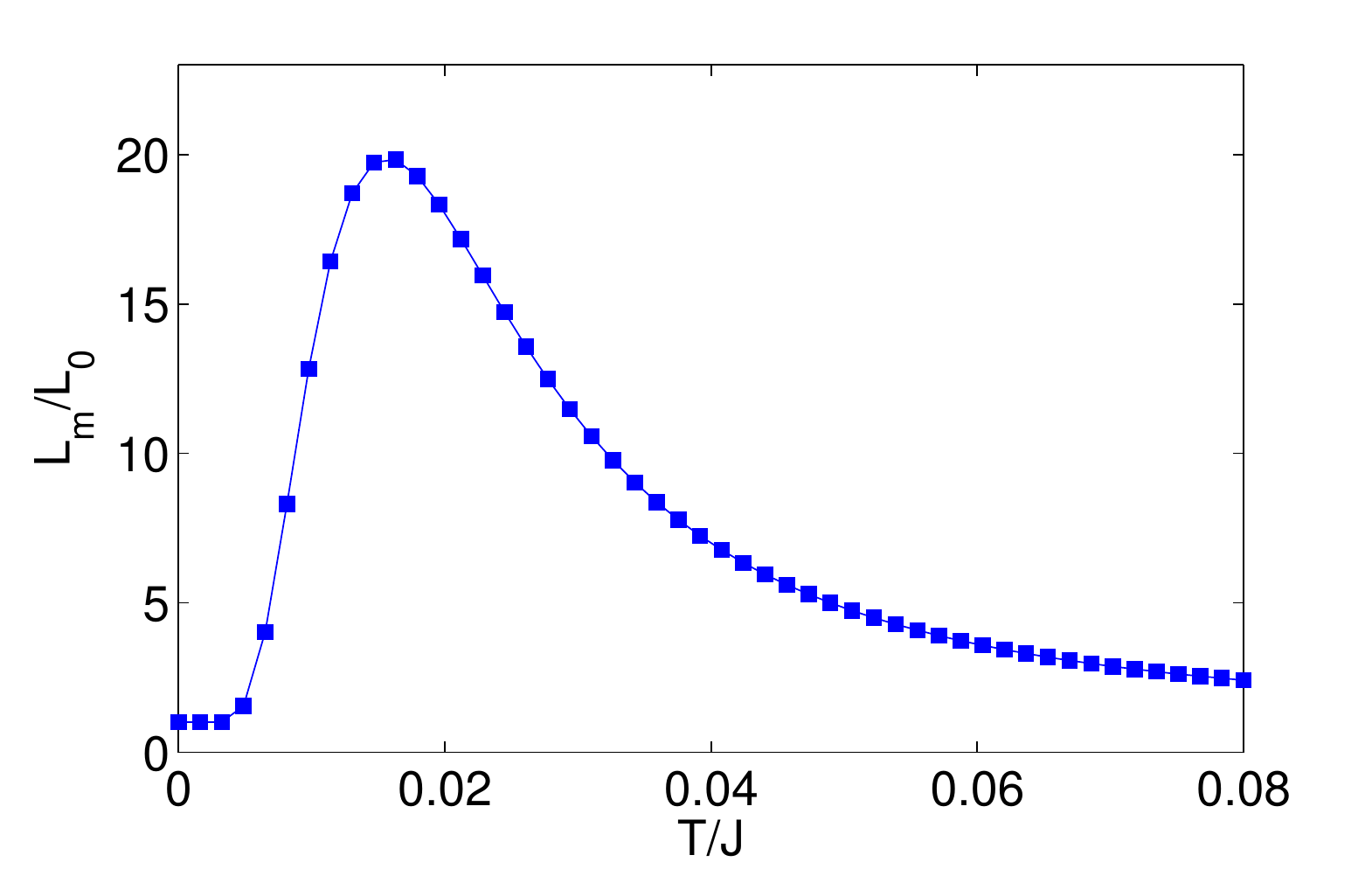} 
  \caption{[Color online] The estimation of the temperature dependent Lorentz number for the DKSL at the doping $x=0.1$.} \label{fig:l_m}
\end{figure}

The DKSL has well-defined low energy excitations and satisfies the Luttinger's theorem. It is a Fermi liquid. However, it has some unusual properties due to the existence of the neutral spinon KSL in the background.  The DKSL has the specific heat coefficient $\gamma=\gamma_d+\gamma_f$, with the dopon specific heat coefficient
\begin{eqnarray}
  \gamma_d=\frac{1}{4}\sum_{\mathscr{A}\mathbf{k}\sigma}(\beta\epsilon_{\mathscr{A}\mathbf{k}}^d)^2\beta\text{sech}^2(\beta\epsilon_{\mathscr{A}\mathbf{k}}^d/2)
  \label{eq:gammad}
\end{eqnarray}
and the spinon sector has the mean field specific heat coefficient
\begin{eqnarray}
  \gamma_f=\frac{1}{2}\sideset{}{'}\sum_{\mathscr{A}\mathbf{k}\sigma}(\beta\epsilon_{\mathscr{A}\mathbf{k}\sigma}^f)^2\beta\text{sech}^2(\beta\epsilon_{\mathscr{A}\mathbf{k}\sigma}^f/2)
  \label{eq:gamma}
\end{eqnarray}
The specific heat coefficient in the DKSL is no longer temperature independent at low temperatures. In the transport measurements, the spinon sector has the contribution in the thermal conductivity, but no  electric conductivity due to its charge neutrality. Therefore, the Wiedemann-Franz law breaks down. We can estimate the Lorentz number of the DKSL as follows
\begin{eqnarray}
  L_m=\frac{\kappa}{T\sigma}\sim\frac{\gamma_d+\gamma_f}{\gamma_d}L_0
  \label{eq:lm}
\end{eqnarray}
with $L_0\equiv\frac{\gamma_d}{T\sigma}=\frac{\pi^2}{3}\left( \frac{k_B}{e} \right)^2$. Here we assume that the mean free paths for the dopon and spinon are close to each other. More detail of the mean field path is beyond the scope of this paper and left to further investigation.  For $x=0.1$, the temperature dependent Lorentz number is shown in Fig. \ref{fig:l_m}. There is a maximum at the temperature around $T^*=0.31E_\text{vp}^\text{eff}$.

The spin susceptibility of the DKSL is given as $\chi=\chi_d+\chi_f$, with the dopon spin susceptibility
\begin{eqnarray}
  \chi_d=\frac{1}{16}\sum_{\mathscr{A}\mathbf{k}\sigma}\beta\text{sech}^2(\beta\epsilon_{\mathscr{A}\mathbf{k}}^d/2)
  \label{eq:chi_d}
\end{eqnarray}
which has the temperature independent Pauli behavior at low temperature and is proportional to the density of states in the dopon sector. The calculation of the spinon spin susceptibility $\chi_f$ is a little tricky\footnote{See the supplementary material for the detail.}. The spinon sector is a $p$-wave superconductor and has vanished density of states at zero temperature, however, the magnetic spin response is finite when the magnetic field is along the $z$ direction. The mean field spin susceptibility in the spinon sector  is given as
\begin{eqnarray}
  \chi_f=\frac{\alpha_\uparrow\alpha_\downarrow}{\alpha_\uparrow+\alpha_\downarrow}
  \label{eq:chi_f0}
\end{eqnarray}
with the definition
\begin{eqnarray}
  \alpha_{\mathbf{k}\uparrow}&=&\frac{1}{\epsilon_{1\mathbf{k}\uparrow}^f\epsilon_{2\mathbf{k}\uparrow}^f}[\epsilon_{1\mathbf{k}\uparrow}^f+\epsilon_{2\mathbf{k}\uparrow}^f-\frac{4g_-(\mathbf{k})g_-(-\mathbf{k})}{\epsilon_{1\mathbf{k}\uparrow}^f+\epsilon_{2\mathbf{k}\uparrow}^f}],\nonumber\\
  \alpha_{\mathbf{k}\downarrow}&=&\frac{1}{\epsilon_{1\mathbf{k}\downarrow}^f\epsilon_{2\mathbf{k}\downarrow}^f}[\epsilon_{1\mathbf{k}\downarrow}^f+\epsilon_{2\mathbf{k}\downarrow}^f-\frac{4h_+(\mathbf{k})h_+(-\mathbf{k})}{\epsilon_{1\mathbf{k}\downarrow}^f+\epsilon_{2\mathbf{k}\downarrow}^f}],\nonumber\\
  \alpha_\sigma&=&\sideset{}{'}\sum_{\mathbf{k}}\alpha_{\mathbf{k}\sigma}.
  \label{eq:alpha_k0}
\end{eqnarray}
At finite doping $x$, the spinon spin susceptibility is given as
\begin{eqnarray}
  \chi_f(x)=\frac{J}{J_\text{eff}}\chi_f(0)
  \label{eq:chifx}
\end{eqnarray}
For finite doping, the effective exchange constant $J_\text{eff}$ is reduced, however, the spinon magnetic response will be enhanced. This enhancement is also observed in the numerical results for the Kitaev model with the spin vacancy\cite{Trousselet2011}. 

At zero temperature, the spinon sector has a vanished specific heat coefficient, but finite spin susceptibility. So it has a the Wilson ration 
\begin{eqnarray}
  R=\frac{\chi_d+\chi_f}{\gamma_d+\gamma_f}
  \label{eq:R}
\end{eqnarray}
larger than 1 on the mean field level. For $x=0.1$, $R\simeq8$.

In this paper, we study the doped Kitaev spin liquid which has the potential experimental realization in the doped layered iridates.  We calculate the electron Green's function and the thermodynamic properties for the DKSL which can be measured in the further experiments.  We find the DKSL to be a Fermi liquid state but with a temperature dependent Lorentz number and a large Wilson ratio. 

JWM thanks T. M. Rice, F. C. Zhang and W. Q. Chen for useful discussions. The work is supported by Swiss National Fonds.  

\bibliography{DKSL.bib}
\appendix
\section{Supplementary material}
\subsection{The spinon susceptibility}
In the following, we will calculate $\chi_f$ at zero temperature. To calculate the spinon magnetic spin response, we add the Zeeman term to the Kitaev bilinear model
\begin{eqnarray}
  H_f[h]=H_f+\frac{1}{2}\sum_{\mathbf{k}\sigma}\psi_{\mathbf{k}\sigma}^\dag (\mu_f+\sigma\frac{h}{2})M_h\psi_{\mathbf{k}\sigma}
  \label{eq:mh}
\end{eqnarray}
with $M_h=\text{diag}(1,1,-1,-1)$ and the chemical potential $\mu_f$ is to enforce single-occupancy constraint on the mean field level. At $h=0$, $\mu_f=0$. The spinon sector now has the $h$-dependent eigenvalues $\pm\epsilon_{1\mathbf{k}\uparrow}^f(h)$, $\pm\epsilon_{2\mathbf{k}\uparrow}^f(h)$, $\epsilon_{1\mathbf{k}\downarrow}^f(h)$ and $\epsilon_{2\mathbf{k}\downarrow}^f(h)$ with the definition
\begin{eqnarray}
  \epsilon_{1\mathbf{k}\uparrow}^f(h)&=&\sqrt{\rho_{\mathbf{k}}+\sqrt{\tau_{\mathbf{k}}}},\quad
   \epsilon_{2\mathbf{k}\uparrow}^f(h)=\sqrt{\rho_{\mathbf{k}}-\sqrt{\tau_{\mathbf{k}}}},\nonumber\\
   \epsilon_{1\mathbf{k}\downarrow}^f(h)&=&\sqrt{r_{\mathbf{k}}+\sqrt{t_{\mathbf{k}}}},\quad
   \epsilon_{2\mathbf{k}\downarrow}^f(h)=\sqrt{r_{\mathbf{k}}+\sqrt{t_{\mathbf{k}}}}\nonumber
   \label{eq:ukh}
\end{eqnarray}
where
\begin{eqnarray}
  \rho_{\mathbf{k}}&=&\frac{1}{2}((\epsilon_{1\mathbf{k}\uparrow}^f)^2+(\epsilon_{2\mathbf{k}\uparrow}^f)^2)+(\mu_f+\frac{h}{2})^2,\nonumber\\
  \tau_{\mathbf{k}}&=&( \frac{(\epsilon_{1\mathbf{k}\uparrow}^f)^2-(\epsilon_{2\mathbf{k}\uparrow}^f)^2}{2} )^2+4g_-(\mathbf{k})g_-(-\mathbf{k})(\mu_f+\frac{h}{2})^2,\nonumber\\
  r_{\mathbf{k}}&=&\frac{1}{2}((\epsilon_{1\mathbf{k}\downarrow}^f)^2+(\epsilon_{2\mathbf{k}\downarrow}^f)^2)+(\mu_f-\frac{h}{2})^2,\nonumber\\
  t_{\mathbf{k}}&=&(\frac{(\epsilon_{1\mathbf{k}\downarrow}^f)^2-(\epsilon_{2\mathbf{k}\downarrow}^f)^2}{2})^2+4h_+(\mathbf{k})h_+(-\mathbf{k})(\mu_f-\frac{h}{2})^2\nonumber
  \label{eq:rt}
\end{eqnarray}
The mean field free energy can be obtained from the mean field Hamiltonian
\begin{eqnarray}
  F_f(h)&=&-\frac{1}{\beta}\text{tr}(-\beta H_f(h))\nonumber\\
  &=&-\frac{1}{\beta}\sideset{}{'}\sum_{\mathscr{A}\mathbf{k}\sigma}\ln(1+\cosh[\beta \epsilon_{\mathscr{A}\mathbf{k}\sigma}^f(h)])
  \label{eq:free_mf1}
\end{eqnarray}
The field dependent chemical potential is determined by
\begin{eqnarray}
  \frac{\partial F_f(h)}{\partial \mu_f}=\sideset{}{'}\sum_{\mathscr{A}\mathbf{k}\sigma}\tanh(\beta\epsilon_{\mathscr{A}\mathbf{k}\sigma}^f(h)/2)\frac{\partial \epsilon_{\mathcal{A}\mathbf{k}\sigma}^f(h)}{\partial \mu_f}=0
  \label{eq:cp}
\end{eqnarray}
At zero temperature, $\tanh(\beta\epsilon)=1$. We take the partial derivative of Eq. (\ref{eq:cp} ) with respect to magnetic field h and then we have the following relation at $h=0$
\begin{eqnarray}
  \sideset{}{'}\sum_{\mathscr{A}\mathbf{k}\sigma}\frac{\partial^2\epsilon_{\mathscr{A}\mathbf{k}\sigma}^f}{\partial\mu_f\partial h}=\sideset{}{'}\sum_{\mathbf{k}\sigma}(\frac{\partial\mu_f}{\partial h}+\sigma\frac{1}{2})\alpha_{\mathbf{k}\sigma}=0
  \label{eq:muf_h}
\end{eqnarray}
with the definition
\begin{eqnarray}
  \alpha_{\mathbf{k}\uparrow}&=&\frac{1}{\epsilon_{1\mathbf{k}\uparrow}^f\epsilon_{2\mathbf{k}\uparrow}^f}[\epsilon_{1\mathbf{k}\uparrow}^f+\epsilon_{2\mathbf{k}\uparrow}^f-\frac{4g_-(\mathbf{k})g_-(-\mathbf{k})}{\epsilon_{1\mathbf{k}\uparrow}^f+\epsilon_{2\mathbf{k}\uparrow}^f}],\nonumber\\
  \alpha_{\mathbf{k}\downarrow}&=&\frac{1}{\epsilon_{1\mathbf{k}\downarrow}^f\epsilon_{2\mathbf{k}\downarrow}^f}[\epsilon_{1\mathbf{k}\downarrow}^f+\epsilon_{2\mathbf{k}\downarrow}^f-\frac{4h_+(\mathbf{k})h_+(-\mathbf{k})}{\epsilon_{1\mathbf{k}\downarrow}^f+\epsilon_{2\mathbf{k}\downarrow}^f}]  
  \label{eq:alpha_k}
\end{eqnarray}
Thus at $h=0$, we have
\begin{eqnarray}
  \frac{\partial \mu_f}{\partial h}=-\frac{1}{2}\frac{\alpha_{\uparrow}-\alpha_{\downarrow}}{\alpha_{\uparrow}+\alpha_{\downarrow}},\quad \alpha_{\sigma}=\sideset{}{'}\sum_{\mathbf{k}}\alpha_{\mathbf{k}\sigma}
  \label{eq:cpp}
\end{eqnarray}
\begin{figure}[b]
  \includegraphics[width=6.5cm]{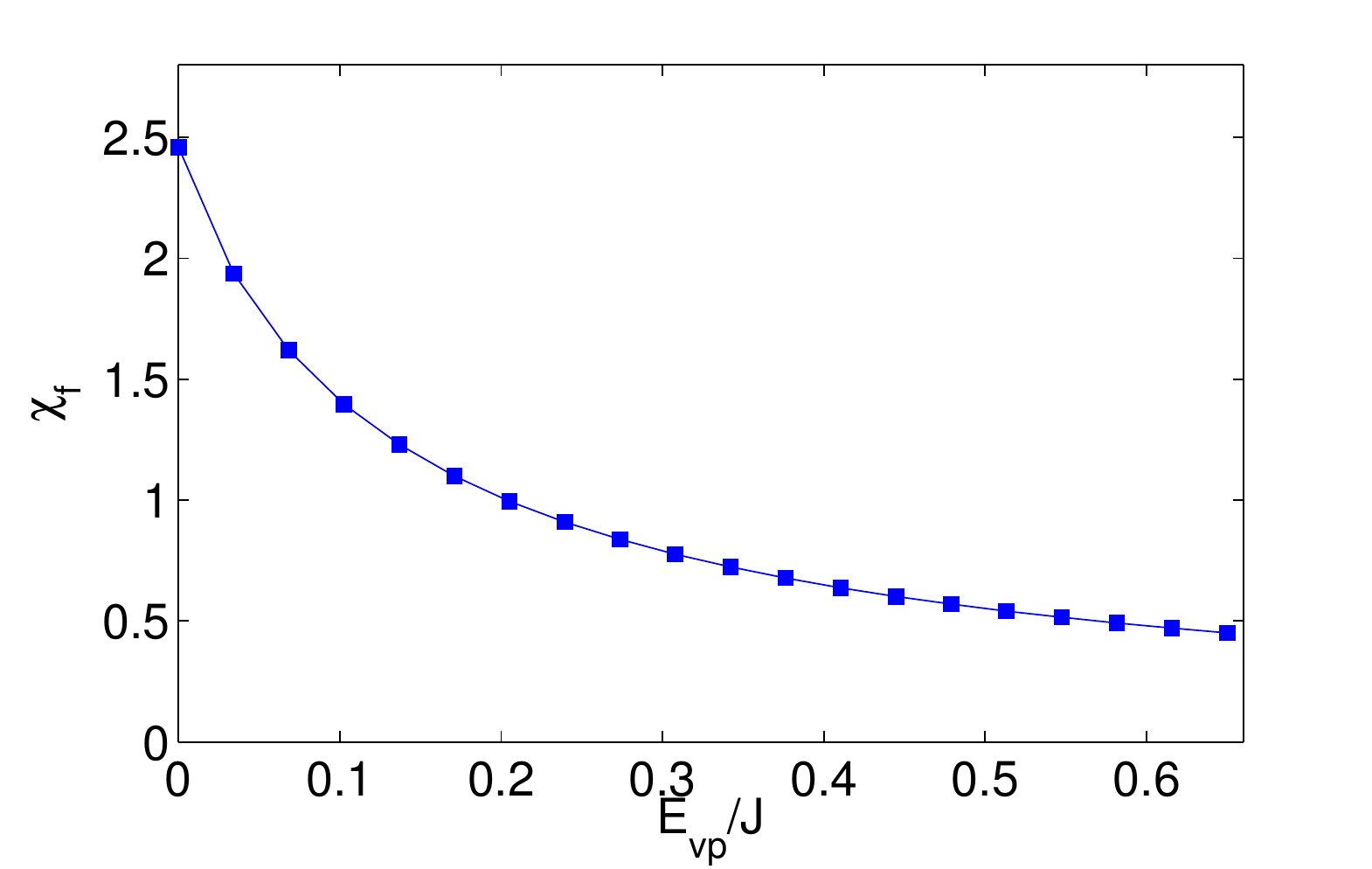} 
  \caption{The spin susceptibility at $T=0$.} \label{fig:chi_Ev}
\end{figure}
The spin susceptibility is given as
\begin{eqnarray}
  \chi=-\frac{\partial^2 F_f}{\partial h^2}&=&\sideset{}{'}\sum_{\mathscr{A}\mathbf{k}\sigma}\left(\frac{1}{2}\frac{\partial\epsilon_{\mathscr{A}\mathbf{k}\sigma}^f}{\partial h}\beta\text{sech}(\beta\epsilon_{\mathscr{A}\mathbf{k}\sigma}^f)^2\right.\nonumber\\
  &&+\left.\frac{\partial^2\epsilon_{\mathscr{A}\mathbf{k}\sigma}^f}{\partial h^2}\tanh(\beta \epsilon_{\mathscr{A}\mathbf{k}\sigma}^f/2)\right.) 
  \label{eq:chi_k}
\end{eqnarray}
At $h=0$, we have $\frac{\partial\epsilon_{A\mathbf{k}\sigma}}{\partial h}=0$ and the spinon mean field spin susceptibility  at zero temperature
\begin{eqnarray}
  \chi_f=(\frac{\partial \mu_f}{\partial h}+\frac{1}{2})^2\alpha_{\uparrow}+(\frac{\partial \mu_f}{\partial h}-\frac{1}{2})^2\alpha_{\downarrow}=\frac{\alpha_\uparrow\alpha_\downarrow}{\alpha_\uparrow+\alpha_\downarrow}
  \label{eq:chi_f}
\end{eqnarray}

Also the spinon spin susceptibility $\chi_f$ is sensitive to the nearest neighbor vortex pair energy $E_\text{vp}$. Without doping, $x=0$, the $E_\text{vp}$-dependent spinon spin susceptibility $\chi_f$ is shown in Fig. \ref{fig:chi_Ev}. 
\end{document}